\documentclass[aps,prl,reprint,superscriptaddress,nofootinbib,floatfix]{revtex4-1}

\usepackage{amsmath,amssymb,bm}
\usepackage{graphicx}
\usepackage{booktabs}
\usepackage{balance}
\usepackage{hyperref}
\hypersetup{hidelinks}

\newcommand{\ER}{E_{\rm R}}
\newcommand{\vmin}{v_{\min}}
\newcommand{\dd}{\mathrm{d}}
\newcommand{\Nsig}{N_{\rm sig}}
\newcommand{\Nlow}{N_{\rm low}}
\newcommand{\Nhi}{N_{>271}}

\begin{document}

\title{Model-Independent Sideband Constraints on Inelastic Dark Matter
       at the LZ High-Recoil Candidate}

\author{David Delepine}
\email{delepine@ugto.mx}
\affiliation{Departamento de F\'isica, Universidad de Guanajuato,
Loma del Bosque 103, 37150 Le\'on, Guanajuato, Mexico}
\author{Shaaban Khalil}
\email{skhalil@zewailcity.edu.eg}
\affiliation{Center for Fundamental Physics, Zewail City of Science and
Technology, Sheikh Zayed, 12588, Giza, Egypt}

\date{\today}

\begin{abstract}
The LUX-ZEPLIN experiment has reported one nuclear-recoil event at
$248\pm23_{\rm stat}\pm23_{\rm sys}\,{\rm keV}$ in a $2.84\,{\rm tonne\text{-}year}$
exposure, with a local significance of $3.4\sigma$.
%Inelastic dark matter is a natural candidate because the kinematic threshold suppresses low-speed recoils.
Inelastic (endothermic) dark matter is a natural candidate because its kinematic threshold suppresses low-energy recoils and shifts the signal toward higher recoil energies.
 We perform a model-independent analysis of the LZ data using two sidebands, without specifying an ultraviolet completion or a particular cross-section parametrization. Our results therefore apply broadly to inelastic dark matter scenarios with coherent spin-independent nuclear scattering.
%We perform a model-independent, two-sided sideband analysis of the LZ data
%without assuming any ultraviolet completion or cross-section ansatz, so that the resulting constraints apply to any inelastic dark matter model with coherent spin-independent scattering.
For each $(m_\chi,\delta)$ we compute $\Nlow/\Nsig$ and $\Nhi/\Nsig$---the
predicted count ratios in the low-energy [$14$, $225$\,keV) and high-energy
($271$, $800$\,keV) sidebands relative to the signal region [$225$, $271$\,keV],
normalised so that $\Nsig=1$.
%These two ratios vary in opposite directions with $\delta$: there exists a unique
%balanced splitting $\delta^\star$ at which $\Nlow/\Nsig = \Nhi/\Nsig \equiv N^\star$,
%minimising the worst-case single-sideband prediction $\max(\Nlow/\Nsig,\,\Nhi/\Nsig)$.
%For $m_\chi=1\,{\rm TeV}$ we find $\delta^\star=339\,{\rm keV}$ and
%$N^\star=1.77$, corresponding to a joint Poisson significance of ${\sim}1.9\sigma$.
%Across the mass range $500\,{\rm GeV}$--$3\,{\rm TeV}$, the tension at $\delta^\star$
%ranges from negligible (${\lesssim}1.3\sigma$ below $700\,{\rm GeV}$)
%to ${\sim}2.5\sigma$ at $3\,{\rm TeV}$.
%An independent nuclear-response constraint arises from the Helm form factor:
%at $248\,{\rm keV}$ the coherent xenon response is
%$|F_{\rm Xe}^{\rm Helm}|^2\approx3\times10^{-4}$, near its second diffraction
%zero at ${\sim}284\,{\rm keV}$.
%A targeted LZ search in the interval $(271,800)\,{\rm keV}$ is identified as the
%decisive test of all inelastic interpretations with $\delta\gtrsim330\,{\rm keV}$.
 The two sideband-to-signal ratios vary in opposite directions with \(\delta\) and are simultaneously optimized at the unique splitting \(\delta^\star\) where they become equal.
For \(m_\chi=1\,{\rm TeV}\), we obtain \(\delta^\star=339\,{\rm keV}\) and \(N^\star=1.77\), corresponding to a combined Poisson tension of approximately \(1.9\sigma\). Over the mass range \(m_\chi=500\,{\rm GeV}\)–\(3\,{\rm TeV}\), the tension evaluated at \(\delta^\star\) increases from \(\lesssim1.3\sigma\) for \(m_\chi\lesssim700\,{\rm GeV}\) to approximately \(2.5\sigma\) at \(m_\chi=3\,{\rm TeV}\). Additional suppression arises from the xenon nuclear response: at \(E_R=248\,{\rm keV}\), the Helm form factor is \(|F_{\rm Xe}^{\rm Helm}|^2\simeq3\times10^{-4}\), close to its second diffraction zero near \(E_R\simeq284\,{\rm keV}\). A dedicated LZ search over \(271<E_R<800\,{\rm keV}\) would therefore provide a decisive test of inelastic dark matter scenarios with \(\delta\gtrsim330\,{\rm keV}\), under the assumed coherent spin-independent interaction and halo model.

\end{abstract}

\maketitle

\emph{Introduction.}---The LUX-ZEPLIN (LZ) Collaboration recently reported one
nuclear-recoil (NR) event at $\ER=248\pm23_{\rm stat}\pm23_{\rm sys}\,{\rm keV}$
in a region of very low known background, with local significance $3.4\sigma$ and
global significance $2.6\sigma$~\cite{LZ}.  Although the event is not evidence for
discovery, its unusually high recoil energy motivates inelastic dark-matter
(DM) scattering,
\begin{equation}
 \chi_1+N\longrightarrow\chi_2+N, \qquad \delta=m_{\chi_2}-m_{\chi_1}>0,
 \label{eq:process}
\end{equation}
because the kinematic threshold $\delta$ preferentially suppresses the large
low-energy recoil population that elastic DM would produce~\cite{TSW}.
%The event has generated a large number of DM interpretations, spanning
%Higgsino and neutralino models~\cite{Freese,FanReece,DuWang,BisDM,CheungKang,BisonGNMSSM,ChatRad,Rodd},
%inelastic singlet-doublet and scalar models~\cite{Su2025,DiMauroKE,WangXiao,BorahSDFerm,BandSDScal,LeeYoun,KumarUBL,LianYang},
%gauge-mediated and other extensions~\cite{AhmedLeontaris,KumarBL,MahaPaul,FanHe,DasNom},
%exothermic scenarios~\cite{deLima,BaerBarger},
%axion- and ALP-portal models~\cite{Unwin,AnGao},
%solar-capture tests~\cite{DiMauroSolar},
%and boosted-DM interpretations~\cite{FanHeBoosted,HeiZim}.
The LZ excess has prompted a broad range of dark-matter interpretations, including Higgsino and neutralino scenarios~\cite{Freese,FanReece,DuWang,BisDM,CheungKang,BisonGNMSSM,ChatRad,Rodd}; inelastic singlet–doublet and scalar models~\cite{Su2025,DiMauroKE,WangXiao,BorahSDFerm,BandSDScal,LeeYoun,KumarUBL,LianYang}; gauge-mediated and other extensions~\cite{AhmedLeontaris,KumarBL,MahaPaul,FanHe,DasNom}; exothermic scenarios~\cite{deLima,BaerBarger}; axion- and axion-like-particle portal models~\cite{Unwin,AnGao}; solar-capture probes~\cite{DiMauroSolar}; and boosted-dark-matter interpretations~\cite{FanHeBoosted,HeiZim}.

%Reference~\cite{Rodd} demonstrates the sideband tension for the higgsino
%by explicit calculation of its $Z$-exchange cross section, and finds it persists
%over a range of non-thermal masses ($500\,{\rm GeV}$--$100\,{\rm TeV}$).
%Our sideband ratios (Eq.~\ref{eq:ratios}) are constructed so that the
%model-dependent cross section $\sigma_p$ cancels analytically---the only
%inputs are kinematics $(m_\chi,\delta)$, the Standard Halo Model (SHM) velocity distribution, and the
%coherent nuclear response. This makes our constraints independent of the
%coupling structure ($Z$-exchange, scalar mediator, contact operator, etc.),
%extending the sideband argument from a single interaction hypothesis scanned
%over mass to the full class of coherent spin-independent inelastic models.

%We make no assumption about the ultraviolet completion, the DM production mechanism,
%or the astrophysical cross section.  Our  inputs are the LZ exposure,
%the SHM velocity distribution, and the Helm form factor
%for coherent spin-independent (SI) scattering---the minimal kinematic and
%nuclear-structure ingredients shared by all inelastic SI models.

Reference~\cite{Rodd} demonstrated the sideband tension for an inelastic
Higgsino by explicitly calculating the $Z$-mediated scattering cross section,
finding that the tension persists over the  mass range
$500\,{\rm GeV}\lesssim m_\chi\lesssim100\,{\rm TeV}$. Here, we
formulate the analysis in terms of the sideband-to-signal ratios defined in
Eq.~\eqref{eq:ratios}. In these ratios, the overall scattering normalization
$\sigma_p$ cancels analytically, leaving only the spectral dependence on the
inelastic kinematics $(m_\chi,\delta)$, the Standard Halo Model (SHM) velocity
distribution, and the coherent xenon nuclear response. Our results therefore
extend the sideband argument beyond a specific $Z$-exchange realization to the
broader class of inelastic dark-matter models whose scattering is described
by the same coherent spin-independent nuclear response with a factorized,
momentum-independent normalization.

The analysis does not require a specific ultraviolet completion or dark-matter
production mechanism. Its physical inputs are the SHM velocity distribution
and the Helm form factor, together with the LZ recoil-energy intervals and
detector response. The resulting constraints are therefore independent of the
overall normalization proportional to $\rho_\chi\sigma_p$, but remain
conditional on the assumed halo model, coherent spin-independent interaction,
and recoil-spectrum dependence.

%A recent Higgsino benchmark~\cite{Freese} quotes a fixed reference cross section
%$\sigma_{\rm fixed}=1.86\times10^{-39}\,{\rm cm}^2$.  Rather than adopting
%Our central questions are: what cross section is \emph{required} to explain the LZ
%event?  How many events does the same model predict in the low-energy and
%high-energy sidebands?  The tension between these sideband predictions and the LZ
%observations quantifies the model-independent challenge for any inelastic SI interpretation.

\emph{Kinematic setup.}---For nucleus mass $m_N$ and DM--nucleus reduced mass
$\mu_{\chi N}=m_\chi m_N/(m_\chi+m_N)$, the minimum speed for  a
recoil $\ER$ is given as
\begin{equation}
 \vmin(\ER)=\frac{1}{\sqrt{2m_N\ER}}
 \left(\frac{m_N\ER}{\mu_{\chi N}}+\delta\right).
 \label{eq:vmin}
\end{equation}
This is minimised at the \emph{optimal recoil}
\begin{equation}
 E_R^\star=\frac{\mu_{\chi N}}{m_N}\,\delta,
 \qquad
 \vmin^\star=\sqrt{\frac{2\delta}{\mu_{\chi N}}},
 \label{eq:opt}
\end{equation}
where the halo integral is largest.  The rate per unit target mass is
\begin{equation}
 \frac{\dd R}{\dd\ER}=\frac{\rho_\chi}{m_\chi}
 \sum_T\frac{\xi_T}{m_T}\,
 \frac{m_T\,\sigma_T^0}{2\mu_{\chi T}^2}\,
 |F_T(q)|^2\,\eta[\vmin(\ER)],
 \label{eq:rate}
\end{equation}
Here $\rho_\chi=0.3\,{\rm GeV\,cm^{-3}}$ is the local DM density; $\xi_T$ and
$m_T\simeq A_T\times0.931\,{\rm GeV}$ are the abundance and isotope $T$ mass;
$\mu_{\chi T}=m_\chi m_T/(m_\chi+m_T)$ is the DM--nucleus reduced mass and
$q=\sqrt{2m_T\ER}$ is the momentum transfer; and
\begin{equation*}
 \sigma_T^0=\sigma_p\,A_T^2\!\left(\frac{\mu_{\chi T}}{\mu_{\chi p}}\right)^{\!2}
\end{equation*}
is the coherent SI cross section on nucleus $T$ (proportional to $A_T^2$; it
cancels in every sideband ratio).
The halo integral $\eta(v)=\int_{|\bm{u}|>v}\dd^3u\,f_{\rm lab}(\bm{u})/|\bm{u}|$
is the mean inverse speed, where $\bm{u}$ is the DM velocity in the Earth's rest
frame and $f_{\rm lab}(\bm{u})$ is the lab-frame velocity distribution.
We adopt the SHM truncated Maxwellian with
$v_0=238$, $v_E=232$, $v_{\rm esc}=544\,{\rm km\,s^{-1}}$~\cite{LewinSmith}.

\begin{figure}[htb!]
 \centering
 \includegraphics[width=\columnwidth]{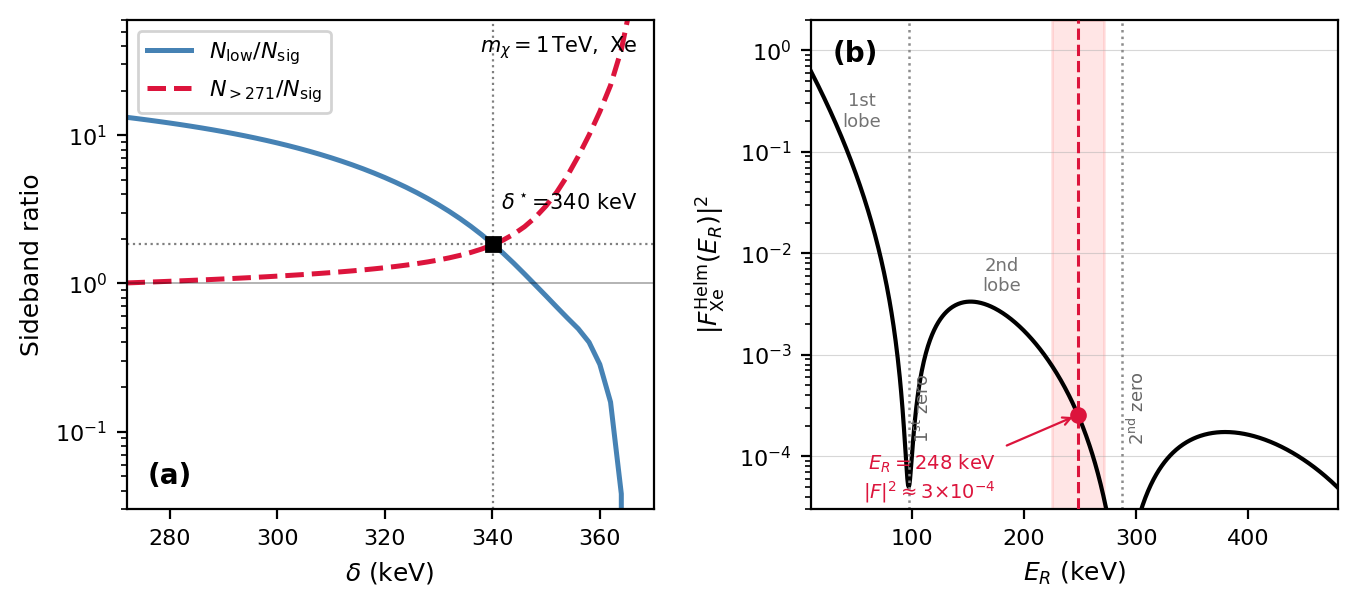}
 \caption{(a) Low- and high-energy sideband ratios vs.\ mass splitting $\delta$
 for $m_\chi=1\,{\rm TeV}$, natural xenon (SHM, coherent SI) with $\Nsig\equiv1$.  The blue solid (red dashed) curve is $\Nlow/\Nsig$
 ($\Nhi/\Nsig$); the crossing at $\delta^\star=339\,{\rm keV}$, $N^\star=1.77$
 (black square) marks the balanced point $\delta^\star$.
 (b) Helm form factor squared for natural xenon (abundance-weighted)
 vs.\ recoil energy.  }
 \label{fig:main}
\end{figure}

For the Helm form factor~\cite{LewinSmith,Duda},
\begin{equation}
 F_{T}^{\rm Helm}(q)=3\,\frac{j_1(qR_{1,T})}{qR_{1,T}}
 \exp\!\left[-\tfrac{1}{2}(qs)^2\right],
 \label{eq:helm}
\end{equation}
with $R_{1,T}=\sqrt{R_T^2-5s^2}$, $R_T=1.2\,A_T^{1/3}\,{\rm fm}$, and
$s=0.9\,{\rm fm}$ and $j_1(x)$ is the first Bessel function. As shown in fig.1b, 
the coherent rate is strongly suppressed at  momentum transfers which correspond to its roots.
For natural xenon (abundance-weighted average over the seven stable isotopes
$A=128$--$136$), we find
\begin{equation}
 |F_{\rm Xe}^{\rm Helm}(248\,{\rm keV})|^2 \approx 3\times10^{-4},
 \label{eq:F2}
\end{equation}
only ${\sim}36\,{\rm keV}$ below the second diffraction zero at ${\sim}284\,{\rm keV}$
(Fig.~\ref{fig:main}b).  This is suppressed by a factor of ${\sim}4000$ relative
to the zero-momentum form factor $F_T(0)=1$.  Any coherent SI model must
accommodate this near-zero nuclear response at the candidate energy.

\emph{Two-sided sideband analysis.}---We define three integration windows in
true recoil energy: the signal region $[225, 271]\,{\rm keV}$ spanning the
published LZ ROI near the candidate; the low sideband $[14,225]\,{\rm keV}$;
and the high sideband $[271, 800]\,{\rm keV}$.  For each $(m_\chi, \delta)$ we
compute
\begin{equation}
 \frac{\Nlow}{\Nsig} \equiv \frac{\int_{14}^{225}\frac{\dd R}{\dd\ER}\,\dd\ER}
                             {\int_{225}^{271}\frac{\dd R}{\dd\ER}\,\dd\ER},
 \qquad
 \frac{\Nhi}{\Nsig} \equiv \frac{\int_{271}^{800}\frac{\dd R}{\dd\ER}\,\dd\ER}
                            {\int_{225}^{271}\frac{\dd R}{\dd\ER}\,\dd\ER},
 \label{eq:ratios}
\end{equation}
where the cross section $\sigma_p$ and exposure $\mathcal{E}$ cancel in each ratio.
We define $\sigma_{\rm req}(m_\chi,\delta)$ as the DM--proton cross section required
to produce the LZ event.
\begin{equation*}
 \mathcal{E}\int_{225}^{271}\frac{\dd R}{\dd\ER}\,\dd\ER\;\bigg|_{\sigma_p=\sigma_{\rm req}} = 1,
\end{equation*}
so that $\Nsig\equiv1$ by construction.  All sideband ratios in Eq.~(\ref{eq:ratios})
are independent of $\sigma_{\rm req}$.
The interpretation is: if an inelastic model with parameters $(m_\chi,\delta)$
is responsible for the single LZ event, it must have $\sigma_p=\sigma_{\rm req}$,
and with this value one can compute  $\Nlow$ events in the low-energy sideband and $\Nhi$ events
in the high-energy sideband.
Throughout this paper uppercase $N$ denotes a \emph{predicted} (expected) count,
a function of $(m_\chi,\delta)$; lowercase $n$ denotes an \emph{observed} count,
a fixed number from the LZ data.

\emph{Lemma (opposite monotonicity):} For fixed $m_\chi$, as $\delta$ increases
from the kinematic threshold, $\Nlow/\Nsig$ decreases monotonically (the threshold
closes the low-energy channel) while $\Nhi/\Nsig$ increases monotonically (the
optimal recoil $E_R^\star$ moves toward and eventually past $271\,{\rm keV}$,
loading events above the signal window). So, a
crossing point $\delta^\star$ at which both ratios are equal can be found.

\textit{Balanced splitting $\delta^\star$:} At $\delta=\delta^\star$,
\begin{equation}
 \frac{\Nlow}{\Nsig}\bigg|_{\delta^\star} = \frac{\Nhi}{\Nsig}\bigg|_{\delta^\star}
 \equiv N^\star.
 \label{eq:balanced}
\end{equation}
This is the unique splitting that minimises the joint two-sided sideband
prediction $\max(\Nlow/\Nsig,\,\Nhi/\Nsig)$.  Any other $\delta$ predicts more
events in at least one sideband.

If LZ observes zero events in both sidebands, the joint Poisson
probability---given $N^\star$ expected events independently in each
sideband---is $p=e^{-N^\star}\cdot e^{-N^\star}=e^{-2N^\star}$, and the
corresponding one-sided Gaussian significance is
\begin{equation*}
 Z=\Phi^{-1}(1-e^{-2N^\star}),
\end{equation*}
where $\Phi(z)=(2\pi)^{-1/2}\int_{-\infty}^{z}e^{-t^2/2}\,\dd t$ is the
standard normal cumulative distribution function.
%For general observed sideband counts the exponent becomes %$N_{\rm low}+N_{>271}$.
%At $\delta^\star$, the total predicted sideband count is $N_{\rm tot}=2N^\star$,
%giving the significance quoted above; any other $\delta$ increases the prediction
%in at least one sideband and raises $N_{\rm tot}$.

\begin{table}[t]
%\caption{Two-sided sideband ratios vs.\ mass splitting for $m_\chi=1\,{\rm TeV}$
%and natural xenon (coherent SI, SHM).  Counts $\Nlow$ and $\Nhi$ are normalised to $\Nsig\equiv1$
%(i.e.\ $\sigma_p=\sigma_{\rm req}(m_\chi,\delta)$ as defined in the text).
%$N_{\rm tot}=\Nlow/\Nsig+\Nhi/\Nsig$
%is the total predicted sideband count; the joint Poisson significance
%$Z=\Phi^{-1}(1-e^{-N_{\rm tot}})$ is evaluated for zero observed events in both
%sidebands simultaneously.  The balanced splitting $\delta^\star=339\,{\rm keV}$
%minimises the worst-case sideband prediction $\max(\Nlow/\Nsig,\,\Nhi/\Nsig)$.}
 \caption{Sideband-to-signal ratios as functions of $\delta$ for
$m_\chi=1\,{\rm TeV}$, assuming natural xenon, coherent SI scattering, and the
SHM with $\Nsig=1$, corresponding to
$\sigma_p=\sigma_{\rm req}$.  The two ratios are equal at the balanced
splitting $\delta^\star=339\,{\rm keV}$.}
\label{tab:delta-scan}
\begin{ruledtabular}
\begin{tabular}{rrrrc}
$\delta$ (keV) & $\Nlow/\Nsig$ & $\Nhi/\Nsig$
  & $N_{\rm tot}/\Nsig$ & $Z$ \\
\hline
280 & 11.83 &  1.03 & 12.86 & $4.6\sigma$ \\
295 &  9.48 &  1.09 & 10.58 & $4.1\sigma$ \\
310 &  6.81 &  1.18 &  7.99 & $3.4\sigma$ \\
325 &  4.04 &  1.34 &  5.38 & $2.6\sigma$ \\
$339^\star$ & 1.77 & 1.77 & 3.54 & $1.9\sigma$ \\
355 &  0.40 &  6.18 &  6.58 & $3.0\sigma$ \\
\end{tabular}
\end{ruledtabular}
\end{table}

Table~\ref{tab:delta-scan} scans $\delta$ at $m_\chi=1\,{\rm TeV}$.
Here $N_{\rm tot}=\Nlow/\Nsig+\Nhi/\Nsig$ is the total predicted sideband count
(normalised to $\Nsig\equiv1$), and $Z=\Phi^{-1}(1-e^{-N_{\rm tot}})$.
For $\delta\lesssim330\,{\rm keV}$, the dominant tension is from the low sideband
($\Nlow/\Nsig>4$), corresponding to ${\gtrsim}2.7\sigma$.  For $\delta\gtrsim355\,{\rm keV}$,
the dominant tension shifts to the high sideband ($\Nhi/\Nsig>6$, ${>}3.0\sigma$).
The unique balanced point at $\delta^\star=339\,{\rm keV}$ yields the minimum
worst-case single-sideband prediction, with $N_{\rm tot}=3.54$ events ($N^\star=1.77$ per sideband), or
$Z\approx1.9\sigma$.  This is substantially
milder than the one-sided low-sideband tension reported in earlier
analyses~\cite{LZ}, which examined only $\Nlow/\Nsig$ without accounting for the
high-energy sideband constraint.
Reference~\cite{Rodd} reports $N_{\rm SB}=3.6$--$10.1$ sideband events for
the higgsino at its best-fit $\delta$ (varying over SHM and form-factor
systematics); our balanced-point value $N^\star=1.77$ is smaller because
we optimise $\delta$ to jointly minimise both sidebands, rather than fixing
$\delta$ at the signal-region maximum-likelihood value---a different
optimisation problem that yields the minimum achievable tension over the
full inelastic parameter space.

Figure~\ref{fig:main}(a) shows the crossing graphically on a logarithmic scale.
The two curves intersect at $\delta^\star=339\,{\rm keV}$; on either side of this
point, one sideband ratio rises steeply.

\begin{table}[b]
\caption{Mass scan: balanced splitting $\delta^\star$, balanced count $N^\star$,
and Poisson significance $Z$ at the balanced point $\delta^\star$ (for zero observed
events) as a function of DM mass.  All results use natural xenon (SHM, coherent SI).}
\label{tab:mass-scan}
\begin{ruledtabular}
\begin{tabular}{rrrr}
$m_\chi$ (GeV) & $\delta^\star$ (keV) & $N^\star$ & $Z(\delta^\star)$ \\
\hline
 500 & 332.5 & $<0.5$ & $<0.3\sigma$ \\
 700 & 336.0 & 1.21 & $1.3\sigma$ \\
1000 & 339.2 & 1.77 & $1.9\sigma$ \\
1500 & 341.9 & 2.17 & $2.2\sigma$ \\
2000 & 343.4 & 2.35 & $2.4\sigma$ \\
3000 & 344.9 & 2.51 & $2.5\sigma$ \\
\end{tabular}
\end{ruledtabular}
\end{table}

Table~\ref{tab:mass-scan} shows the mass scan.  For $m_\chi\lesssim700\,{\rm GeV}$,
the joint balanced tension is below $1.3\sigma$; the LZ event  is
kinematically accessible with no significant sideband penalty.  Above
$1\,{\rm TeV}$, $\delta^\star$ rises slowly and $N^\star$ increases, reaching
$2.51$ events per sideband ($N_{\rm tot}=5.02$, $2.5\sigma$) at $3\,{\rm TeV}$.
%Thus, even in the worst case considered here, the joint two-sided sideband
%tension---which LZ has not yet probed through a targeted high-energy
%search---does not exceed $2.5\sigma$ at our reference SHM values.

\textit{Halo and form-factor systematics.}---The reference values above use
$v_0=238$, $v_E=232$, $v_{\rm esc}=544\,{\rm km\,s^{-1}}$ and the standard
Helm parametrization.  To quantify the sensitivity to the uncertainties of these values, we perform a joint grid scan over
$(v_0,v_{\rm esc},v_E,R_T)$ simultaneously: $v_0\in[220,250]$, $v_{\rm esc}\in[498,608]$,
$v_E\in[227,237]\,{\rm km\,s^{-1}}$~\cite{McMillan,Piffl}, and $R_T=1.2\,A^{1/3}\pm0.1\,{\rm fm}$ \cite{Duda}).  The envelope of this joint scan at $m_\chi=1\,{\rm TeV}$ is:
\begin{itemize}
\item The circular speed $v_0$ (220--250$\,{\rm km\,s^{-1}}$) shifts $\delta^\star$
      by less than $0.2\,{\rm keV}$ and has negligible effect on $N^\star$ when
      $v_{\rm esc}$ and $R_T$ are held fixed.
\item The escape speed $v_{\rm esc}$ (498--608$\,{\rm km\,s^{-1}}$) is the dominant
      halo systematic: holding other parameters at their reference values it shifts
      $\delta^\star$ from $315$ to $373\,{\rm keV}$ and $N^\star$ from $1.1$ to $2.5$.
\item A $\pm0.1\,{\rm fm}$ skin-radius shift moves $N^\star$ from $1.1$ to $3.1$
      at reference halo parameters, because the signal window lies near the Helm
      diffraction zero, the effects of  any small change in $R_T$ are amplified.
\end{itemize}
The joint envelope of the full scan gives $\delta^\star\in[310,\,379]\,{\rm keV}$
and $Z\in[0.6,\,3.5]\sigma$.

\emph{Nuclear-response constraint.}---.  From Eq.~(\ref{eq:F2}), the coherent SI rate at
$248\,{\rm keV}$ is suppressed by a factor $\sim 4000$ relative to the
zero-momentum limit.  This is illustrated in Fig.~\ref{fig:main}(b).
The first diffraction zero occurs near $96\,{\rm keV}$; the first lobe
(with maximum $|F|^2\sim 0.003$) extends from $\sim 96$ to $\sim 284\,{\rm keV}$. The nuclear suppression at $248\,{\rm keV}$ has two consequences.
First, the cross section required to produce $\Nsig=1$ is enhanced by a factor
$\sim4000$ relative to what the same model would need at low recoil energy.
For $m_\chi=1\,{\rm TeV}$ and $\delta=\delta^\star$, we find
$\sigma_{\rm req}\approx2\times10^{-40}\,{\rm cm}^2$. %while the Higgsino
%benchmark of Ref.~\cite{Freese} fixes $\sigma_{\rm fixed}=1.86\times10^{-39}\,{\rm cm}^2$,
%i.e.\ $\sigma_{\rm fixed}/\sigma_{\rm req}\approx9$.  A Higgsino with this
%cross section would predict ${\sim}9$ events in the signal region---not one.
Second, the nuclear suppression means that the same $(m_\chi,\delta)$ would
produce a much larger rate in any nucleus whose form factor does not vanish near
$248\,{\rm keV}$. 

Isotope averaging over the seven stable xenon isotopes ($A=128$--$136$,
natural abundances) softens but does not remove this suppression: the node
positions shift by a few percent across isotopes, so the abundance-weighted
$|F_{\rm Xe}^{\rm Helm}|^2$ shown in Fig.~\ref{fig:main}(b) retains a deep
minimum near $284\,{\rm keV}$.%  A more general spin-independent interaction
%or an operator with a different momentum dependence would modify the form
%factor and could shift or fill the diffraction minimum, but the near-zero value
%at $248\,{\rm keV}$ is a robust feature of the coherent isoscalar channel.

\emph{Implications for the LZ analysis.}---
Any inelastic model that explains the $248\,{\rm keV}$ candidate with
$\delta\gtrsim\delta^\star$ necessarily predicts $\Nhi/\Nsig>N^\star$
events above $271\,{\rm keV}$.  This interval lies above the published LZ
search window and has not yet been analyzed for DM signals.  A null result in $[271, 800]\,{\rm keV}$ would impose a $90\%$ C.L.\ upper
limit on the predicted high-sideband count. 
A detection in this window would instead provide a direct measurement of $\delta$
from the peak recoil energy, constituting strong evidence for inelastic DM.

\begin{figure}[t]
 \centering
 \includegraphics[width=0.8\columnwidth]{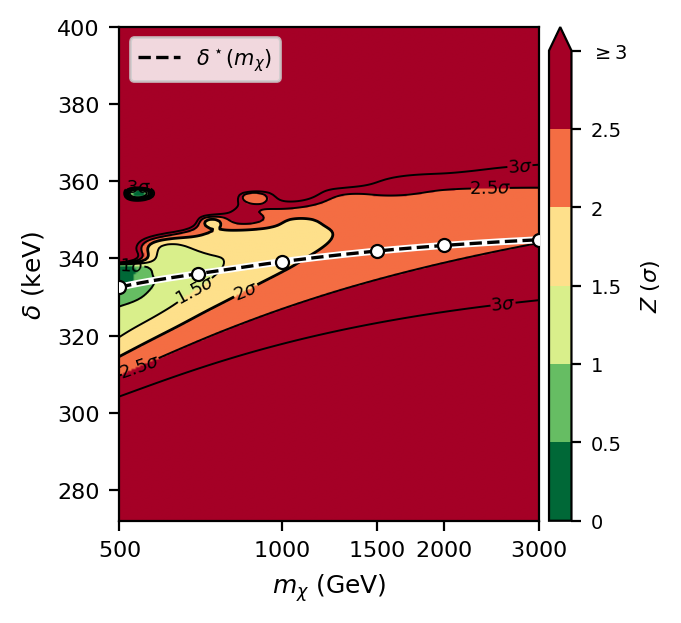}
% \caption{Poisson significance $Z(\delta,m_\chi)=\Phi^{-1}(1-e^{-N_{\rm tot}})$
% in the $(m_\chi,\delta)$ plane (colour scale at right).
% Contour lines are labelled at $Z=1,\,1.5,\,2,\,2.5,\,3\sigma$.
% The white/dashed curve is the balanced ridge $\delta^\star(m_\chi)$
% at which $\max(\Nlow/\Nsig,\,\Nhi/\Nsig)$ is minimised for each $m_\chi$;
% white circles are the reference points from Table~\ref{tab:mass-scan}.
% Warmer colours indicate higher tension; cooler colours indicate lower tension.
% Because $\sigma_p$ cancels analytically in the sideband ratios
% (Eq.~\ref{eq:ratios}), these contours apply identically to any inelastic DM
% model with coherent spin-independent scattering, independently of the
% coupling structure ($Z$-exchange, scalar mediator, contact operator, etc.).}
 \caption{Poisson tension in the $(m_\chi,\delta)$ plane, with contours at
$Z=1,\,1.5,\,2,\,2.5,$ and $3\sigma$. The dashed white curve denotes the
balanced splitting $\delta^\star(m_\chi)$, where the two sideband-to-signal
ratios are equal, and the white circles mark the benchmark points listed in
Table~\ref{tab:mass-scan}. The contours are independent of the overall
cross-section normalization and apply to coherent SI models.}
\label{fig:likelihood2d}
\end{figure}

Given observed counts $n_{\rm low}$ and $n_{>271}$ in the two sidebands with
one event in the signal region ($n_{\rm sig}=1$), the joint likelihood for
inelastic parameters $(\delta,m_\chi)$ is a product of independent Poisson factors,
{\small \begin{equation}
 \mathcal{L}(\delta,m_\chi) =
  \frac{\Nlow^{n_{\rm low}}\,e^{-\Nlow}}{n_{\rm low}!}
  \cdot
  \frac{\Nhi^{n_{>271}}\,e^{-\Nhi}}{n_{>271}!},
 \label{eq:likelihood}
\end{equation}}
where $\Nlow(\delta,m_\chi)$ and $\Nhi(\delta,m_\chi)$ are evaluated at
$\sigma_p=\sigma_{\rm req}(\delta,m_\chi)$.
Because the sideband ratios $\Nlow/\Nsig$ and $\Nhi/\Nsig$ are independent of
$\sigma_p$ (Eq.~\ref{eq:ratios}), the cross section $\sigma_{\rm req}$ is
\emph{uniquely fixed} by the requirement $\Nsig=1$: it is not a free parameter to be
marginalised over, but is determined entirely by $(m_\chi,\delta)$.
The likelihood in Eq.~(\ref{eq:likelihood}) therefore depends only on $(\delta,m_\chi)$,
and its contours in this plane define confidence regions for the inelastic
interpretation directly.
For the current LZ data ($n_{\rm low}=n_{>271}=0$), $\mathcal{L}=e^{-(\Nlow+\Nhi)}$
and the likelihood contours coincide with the isolines of $N_{\rm tot}=\Nlow+\Nhi$,
shown in Fig.~\ref{fig:likelihood2d}. 

\balance

\emph{Conclusion.}---We have presented a model-agnostic analysis of the LZ
high-recoil candidate within the endothermic dark-matter hypothesis, assuming
coherent spin-independent scattering and the Standard Halo Model. For each
$m_\chi$, we identify a balanced splitting $\delta^\star(m_\chi)$ at which the
low- and high-energy sideband-to-signal ratios are equal. Since the overall
normalization $\rho_\chi\sigma_p$ cancels from these ratios, our results do not
depend on a specific ultraviolet completion or production mechanism, but apply
to models sharing the assumed nuclear response and recoil-spectrum dependence.

For $m_\chi=1\,{\rm TeV}$, we find $\delta^\star=339\,{\rm keV}$ and
$1.77$ expected events in each sideband, corresponding to a combined Poisson
tension of approximately $1.9\sigma$ for zero observed events and negligible
background. The interpretation is further challenged by the strong suppression
of the xenon Helm form factor at $E_R=248\,{\rm keV}$,
$|F_{\rm Xe}^{\rm Helm}|^2\simeq3\times10^{-4}$, requiring
$\sigma_{\rm req}\simeq2\times10^{-40}\,{\rm cm}^2$ for
$m_\chi=1\,{\rm TeV}$. A dedicated LZ search over $271<E_R<800\,{\rm keV}$ would directly test this
interpretation. A null result would substantially constrain the relevant
parameter space, whereas an excess with the predicted spectral shape would
support an endothermic dark-matter origin.

\vskip 0.001cm
\textit{Acknowledgements---} The work of D.\,D.\ is supported by Secretaría de Ciencia,
Humanidades, Tecnología e Innovación (SECIHTI) and
Sistema Nacional de Investigadoras e Investigadores
(S.N.I.I.), Mexico.

\end{document}